\documentclass[12pt]{iopart}

\usepackage{iopams}
\usepackage{graphicx}
\usepackage{epstopdf}
\usepackage{bm}
\usepackage{braket}

\newcommand{\hyphen}{\mathchar`-}

\begin{document}

\title[]{Heat transport through a two-level system embedded between two harmonic resonators}

\author{Tsuyoshi Yamamoto${}^{*}$ and Takeo Kato}

\address{Institute for Solid State Physics, the University of Tokyo, Kashiwa, Chiba 277-8581, Japan}
\ead{t.yamamoto@issp.u-tokyo.ac.jp}
\vspace{10pt}
\begin{indented}
\item[]July 2021
\end{indented}

\begin{abstract}
We investigate heat transport through an assembly consisting of a two-level system coupled between two harmonic oscillators, which is described by the quantum Rabi model, as a prototype of nanoscale heat devices using controllable multi-level systems.
Using the noninteracting-blip approximation (NIBA), we find that the linear thermal conductance shows a characteristic temperature dependence with a two-peak structure.
We also show that heat transport is sensitive to model parameters for weak system-bath coupling and strong hybridization between the two-level system and the harmonic oscillators.
This property characteristic of the multi-level system is advantageous for applications such as a heat transistor, and can be examined in superconducting circuits.
\end{abstract}

%
%
%
%
%

\section{Introduction}

Heat transport through nanoscale objects has attracted interest for a long period of time from the viewpoint of both fundamental physics on quantum mechanics~\cite{Schwab2000, Meschke2006, Timofeev2009, Pekola2015} and application to nanoscale heat devices~\cite{Li2012}.
Various heat nanoscale devices (e.g., quantum heat rectification~\cite{Segal2005PRL, Ruokola2009}, quantum heat transistor~\cite{Ojanen2008, Joulain2016}, quantum refrigerator~\cite{Kilgour2018}, and quantum heat engine~\cite{Uzdin2015,Carrega2019,Wiedmann2020}) have been theoretically proposed as analogy to nanoscale electronic devices.
Recently, some nanoscale heat devices have been implemented in various physical systems using nanostructures~\cite{Ronzani2018,Senior2020,Klatzow2019}.

Recent advances in circuit quantum electrodynamics (QED) using superconductor devices~\cite{Xiang2013, Kockum2019, Forn-Diaz2019} have led to considerable developments in the study of heat transport through small quantum objects.
The circuit QED system provides an ideal platform for the study of heat transport~\cite{Ronzani2018, Senior2020}, in which superconducting qubits~\cite{Nakamura1999, Mooij1999, Koch2007} can be coupled to quantum resonators or transmission lines with well-controlled coupling strengths.
In fact, heat transport in superconducting circuits has been measured with high accuracy in recent experiments~\cite{Karimi2020, Maillet2020, Yang2020}.

Heat transport through a small quantum system has been studied theoretically in various models such as a non-linear harmonic oscillator~\cite{Ruokola2009} and a harmonic chain~\cite{Wu2011,Segal2016}.
In particular, a two-level system coupled directly to heat baths modeled as a collection of harmonic oscillators that is called the spin-boson model~\cite{Leggett1987, Weiss_text} has been widely used as a simple prototype system~\cite{Segal2005PRL, Ruokola2011, Saito2013, Segal2014, Yang2014, Taylor2015,Carrega2016, Wang2017, Yamamoto2018NJP, Yamamoto2018PRB}.
In these studies, a specific type of a heat bath called an Ohmic heat bath has been considered.
However, the spin-boson model with the Ohmic heat bath is too simple for the discussion of sophisticated control of nanoscale heat devices, because the number of independent model parameters is not large.
As a model with more parameters, the two-level system embedded between two harmonic oscillators can be considered (see figure~\ref{fig:system}), which is an archetype of small systems with multiple levels described by the quantum Rabi model~\cite{Braak2011}.
Although the quantum Rabi model is still simple and fundamental, it has sufficient flexibility for designing multiple quantum levels of the system.
We should note that heat transport of this setup has indeed been measured in recent experiments using a superconducting circuit~\cite{Ronzani2018, Senior2020}.
However, these implementations are restricted to weak hybridization between the two-level and the harmonic oscillators, and the regime of the ultra-strong hybridization has not been studied yet experimentally and theoretically.

In this work, we explore heat transport through a small quantum system described by the quantum Rabi model between the Ohmic heat baths whose spectral density function is linear to the frequency.
We reveal several features of its heat transport, that cannot be observed in simple systems such as the spin-boson model.
We find that the thermal conductance has a two-peak structure as a function of the temperature when the harmonic oscillators are strongly hybridized with the two-level system under weak system-bath coupling.
In this situation, heat transport becomes sensitive to the parameters (e.g., the natural frequency of the harmonic oscillators and the tunneling splitting of the two-level system), and a certain value of the parameter induces the enhancement of the heat flux.
This sensitivity is advantageous to applications of nanoscale heat devices.
Moreover, we clarify the origin of this sensitivity from the energy spectrum of the quantum Rabi model.

The rest of this paper is organized as follows.
In section~\ref{sec:model}, we introduce the dissipative quantum Rabi model and derive the formula for the linear thermal conductance.
We also present limiting cases, which allow analytic calculation.
In section~\ref{sec:NIBA}, we introduce a non-perturbative approximation, the noninteracting-blip approximation (NIBA)~\cite{Leggett1987, Weiss_text}, and obtain an approximate expression for the linear thermal conductance.
In section~\ref{sec:result}, we discuss how the thermal conductance depends on the temperature and the natural frequency of the harmonic oscillators.
In section~\ref{sec:discussion}, we discuss the experimental feasibility and the rotating wave approximation.
Finally, we summarize our work in section~\ref{sec:summary}.

\section{Formulation}
\label{sec:model}

\begin{figure}[tb]
\centering
\includegraphics[width=0.6\columnwidth]{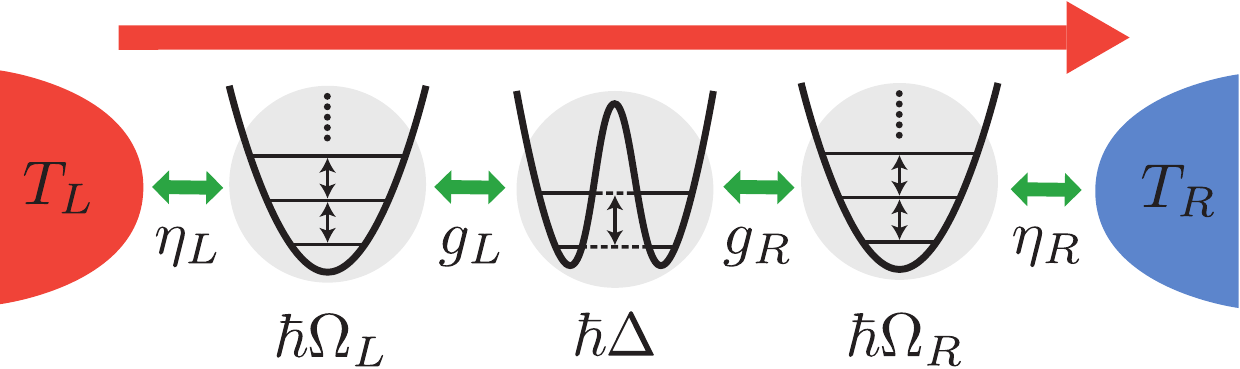}
\caption{\label{fig:system}
Schematic of heat transport though a small system described by the quantum Rabi model, in which a two-level system with a tunneling splitting frequency $\Delta$ is embedded between two harmonic oscillators with a natural frequency $\Omega_r$ ($r=L,R$).
When the temperature of the left heat bath is larger than that of the right one ($T_L>T_R$), heat flow is induced through the small system described by the quantum Rabi model from the left to the right.
In this study, we consider the case of $\Omega_L =\Omega_R \equiv \Omega$ and $\eta_L=\eta_R \equiv \eta$.
}
\end{figure}

In this section, we introduce a model that describes the setup in figure~\ref{fig:system} and formulate a thermal conductance by using the Keldysh formalism, providing the Meir-Wingreen-Landauer-type heat current~\cite{Saito2013}.
In addition, we briefly summarize the weak-coupling theory and renormalization effect due to the system-reservoir coupling.

\subsection{Model}
\label{sec:full-model}

We consider heat transport through a two-level system embedded between harmonic oscillators described by the quantum Rabi model, whose Hamiltonian is given as
\begin{eqnarray}
    \label{eq:Hq}
    & & H_{\rm q} = -\frac{\hbar\Delta}{2}\sigma_x + \sum_{r}\hbar\Omega_r a_r^\dagger a_r + \sum_{r}\hbar g_r\sigma_z\left(a_r+a_r^\dagger\right),
\end{eqnarray}
where $a_r~(a_r^\dagger)$ is an annihilation (a creation) operator of the harmonic oscillator $r$ ($=L,R$), $\sigma_{x,y,z}$ are Pauli matrices, $\Delta$ is a tunneling splitting frequency.
In general, the  two-level system includes a detuning energy, $-\hbar\varepsilon \sigma_z$.
In this paper, we consider the zero-detuning case $(\varepsilon = 0)$ for simplicity.
$\Omega_r$ is a natural frequency of the harmonic oscillator $r$, and the frequency $g_{r}$ represents the coupling strength between the two-level system and the harmonic oscillator $r$ (see figure~\ref{fig:system}).

The two heat baths are modeled as a collection of harmonic oscillators.
The Hamiltonian of the heat bath $r$ $(=L,R)$ is given as
\begin{eqnarray}
    \label{eq:Hbath}
    & & H_{r} = \sum_{k}\hbar\omega_{r,k}b_{r,k}^\dagger b_{r,k}
\end{eqnarray}
with an annihilation (a creation) operator $b_{r,k}$ ($b^\dagger_{r,k}$) of the $k$-th bosonic mode with the frequency $\omega_{r,k}$ in the heat bath $r$.
The total Hamiltonian is given as
\begin{eqnarray}
    & & H = H_{\rm q} + \sum_r (H_r + H_{{\rm int},r}),
\label{eq:Hfull}
\end{eqnarray}
where $H_{{\rm int},r}$ describes the coupling between the harmonic oscillator $r$ and the heat bath $r$:
\begin{eqnarray}
    \label{eq:Hint}
    & & H_{{\rm int},r} = \sum_{k}\hbar\lambda_{r,k}\left(a_r+a_r^\dagger\right)\left(b_{r,k}+b_{r,k}^\dagger\right).
\end{eqnarray}
Here, the frequency $\lambda_{r,k}$ represents the coupling strength between the harmonic oscillator $r$ and the $k$-th bosonic mode in the heat bath $r$.
Each of the heat baths is completely characterized by a spectral density
\begin{eqnarray}
    & & I_r(\omega) = \sum_{k}\lambda_{r,k}^2\delta\left(\omega-\omega_{r,k}\right).
\end{eqnarray}
We assume that the number of bosonic modes is so large that the spectral density $I_r(\omega)$ can be considered as a smooth function of the frequency.
We consider the Ohmic baths, for which the spectral density is linear with respect to $\omega$:
\begin{eqnarray}
    & & I_r(\omega) = 2\eta_r\omega.
\end{eqnarray}
Here, $\eta_r$ represents a dimensionless coupling strength between the heat bath $r$ and the harmonic oscillator $r$.

\subsection{Mapping to the effective spin-boson model}
\label{sec:sb-model}

The dissipative quantum Rabi model~(\ref{eq:Hfull}) can be mapped to the spin-boson model (see the detail derivation in \ref{app:mapping})~\cite{Goorden2004,Zueco2019,Magazzu2019}
\begin{eqnarray}
    \label{eq:Hsb}
    & & H_{\rm SB} 
    = -\frac{\hbar\Delta}{2}\sigma_x
    +\sum_{r,k}
    \hbar\tilde{\omega}_{r,k}\tilde{b}_{r,k}^\dagger \tilde{b}_{r,k} -\frac{\sigma_z}{2}\sum_{r,k}\hbar\tilde{\lambda}_{r,k}
    \left(\tilde{b}_{r,k}+\tilde{b}_{r,k}^\dagger\right),
\end{eqnarray}
where $\tilde{b}_{r,k}$ ($\tilde{b}^\dagger_{r,k}$) is an annihilation (a creation) operator for the $k$-th bosonic mode with the frequency $\tilde{\omega}_{r,k}$ in the effective heat bath $r$ and $\tilde{\lambda}_{r,k}$ represents the coupling strength between the two-level system and the $k$-th bosonic mode in the effective heat bath $r$.
The spectral density characterizing the effective heat baths is expressed by
\begin{eqnarray}
    & & I_{{\rm eff},r}(\omega)
    = \sum_{k}\tilde{\lambda}_{r,k}^2\delta\left(\omega-\tilde{\omega}_{r,k}\right)
    \label{eq:Ieff2}
    = 2\alpha_r\omega\frac{\Omega_r^4}{\left(\Omega_r^2-\omega^2\right)^2+\left(2\Gamma_r\omega\right)^2},
\end{eqnarray}
which is called as the structured spectral density.
Here, the two harmonic oscillators are incorporated into the effective Lorentz-type spectral density, which is characterized by the peak frequency $\Omega_r$, and the peak width $\Gamma_r=\pi\eta_r\Omega_r$, and the dimensionless coupling strength $\alpha_r=\eta_r(4g_r/\Omega_r)^2$.
Note that this mapping is the inverse application of the reaction-coordinate mapping~\cite{Garg1985,Nazir2018}.

\subsection{Thermal conductance}
\label{sec:conductance}

The heat current from the heat bath $r$ into the system described by the quantum Rabi model~(\ref{eq:Hq}) is defined as
\begin{eqnarray}
    \label{eq:current_op}
    & & J_r\equiv-\frac{dH_{r}}{dt}.
\end{eqnarray}
For the effective spin-boson model~(\ref{eq:Hsb}), the steady-state heat current is calculated by the Keldysh formalism as~\cite{Meir1992,Saito2008,Ojanen2008}
\begin{eqnarray}
    & & \Braket{J_r}
    =\frac{\hbar^2}{2}\int_0^\infty d\omega \, \omega I_{{\rm eff},r}(\omega) \left\{{\rm Im}\left[\chi^{R}(\omega)\right]n_r(\omega)-\frac{i}{2}\chi^{<}(\omega)\right\},
\end{eqnarray}
where $n_r(\omega) = (e^
{\beta_r\hbar\omega}-1)^{-1}$ is the Bose-Einstein distribution function for the heat bath $r$, $\beta_r=(k_BT_r)^{-1}$,  $k_{\rm B}$ is the Boltzmann constant, $T_r$ is the temperature of the heat bath $r$, and $\chi^R(t)$ and $\chi^<(t)$ are the retarded and lesser components of the correlation function of the two-level system defined as
\begin{eqnarray}
    & & \chi^R(t)
    =-\frac{i}{\hbar}\theta(t)\Braket{\left[\sigma_z(t),\sigma_z(0)\right]}, \\
    & & \chi^<(t)
    =-\frac{i}{\hbar}\Braket{\sigma_z(t)\sigma_z(0)},
\end{eqnarray}
respectively.

For simplicity, we consider the symmetric case, $\Omega_L=\Omega_R\equiv \Omega$ and $\eta_L=\eta_R \equiv \eta$ hereafter.
Using the conservation law of energy $\braket{J_L}+\braket{J_R}=0$, the heat current can be expressed as
\begin{eqnarray}
    & & \Braket{J_L}
    =\frac{\alpha\gamma\hbar^2}{8}\int_0^\infty d\omega~\omega\tilde{I}_{\rm eff}(\omega){\rm Im}\left[\chi^R(\omega)\right]
    \left[n_L(\omega)-n_R(\omega)\right]
\end{eqnarray}
with $\alpha=\alpha_L+\alpha_R$, $\gamma=4\alpha_L\alpha_R/\alpha^2$, and $\tilde{I}_{\rm eff}(\omega)=\alpha_r^{-1} I_{{\rm eff},r}(\omega)$.
The linear thermal conductance is given as
\begin{eqnarray}
    & & \kappa
    \equiv
    \lim_{T_L\to T_R}\frac{\Braket{J_L}}{T_L-T_R}\\
    & &~~ =\frac{\alpha\gamma\hbar k_B}{8}\! \int_0^\infty\! d\omega\,\tilde{I}_{\rm eff}(\omega){\rm Im}\left[\chi^{R}(\omega)\right]\left[\frac{\beta\hbar\omega/2}{\sinh(\beta\hbar\omega/2)}\right]^2,
\end{eqnarray}
where $T_R=T$ is the temperature and $\beta=(k_BT)^{-1}$ is its inverse temperature.
For the convenience of analysis, we introduce a symmetrized correlation function,
\begin{eqnarray}
    \label{eq:S}
    & & S(t)=\frac{1}{2} \Braket{\sigma_z(t)\sigma_z(0)+\sigma_z(0)\sigma_z(t)}.
\end{eqnarray}
By the fluctuation-dissipation theorem~\cite{Weiss_text}, the imaginary part of the retarded component of the correlation function is related to the Fourier transformation of $S(t)$ as
\begin{eqnarray}
    & & S(\omega)=\hbar\coth\left(\frac{\beta\hbar\omega}{2}\right){\rm Im}\left[\chi^{R}(\omega)\right].
\end{eqnarray}
Then, the linear thermal conductance is rewritten with $S(\omega)$ as
\begin{eqnarray}
    \label{eq:conductance}
    & & \kappa
    =\frac{\alpha\gamma k_B}{16}\int_0^\infty d\omega\,\tilde{I}_{\rm eff}(\omega)S(\omega)\frac{(\beta\hbar\omega)^2}{\sinh(\beta\hbar\omega)}.
\end{eqnarray}
Due to the factor $(\beta \hbar \omega)^2/\sinh(\beta \hbar \omega)$, the thermal conductance is mainly determined by the behavior of $S(\omega)$ in the region of $0< \omega \lesssim k_{\rm B}T/\hbar$.

\subsection{Renormalized Ohmic spin-boson model derived for $\Delta \ll \Omega$}
\label{sec:ohmic}

For $\Delta \ll \Omega$, since the lowest two states of the quantum Rabi model are separated well from the other states, the effective spectral density~(\ref{eq:Ieff2}) can be regarded as a two-level system coupled to the Ohmic bath with a dimensionless system-bath coupling strength $\alpha_r=\eta(4g_r/\Omega)^2$ and a high-frequency cutoff $\Omega$ ($\gg  \Delta$).
Then, heat transport in this regime can be understood by simply referring the results on heat transport via a two-level system~\cite{Yamamoto2018NJP,Saito2013,Ruokola2011,Segal2010,Segal2005PRL}.
This parameter regime is not preferable to control heat transport because this effective two-level system shows a rather simple behavior for heat transport with respect to the parameters.
In fact, the parameters regarding the harmonic oscillators, i.e., $\Omega_r$, $g_r$, and $\eta$, are absorbed into a single parameter $\Delta_*$, which is the renormalized tunneling splitting frequency obtained by the adiabatic renormalization~\cite{Leggett1987}. 
In this study, we focus on the opposite regime, i.e., the case of $\Delta \gtrsim \Omega$, in which the present model exhibits non-trivial transport properties.

\subsection{Weak-coupling regime $(\eta \ll 1)$}
\label{sec:weak_diss}

It is helpful to consider the weak-coupling regime ($\eta\ll 1$).
For the isolated quantum Rabi model ($\eta = 0$), we denote the eigen energy and the corresponding eigen state as $\hbar\omega_n$ and $\ket{n}$, respectively:
\begin{eqnarray}
H_q\ket{n}=\hbar\omega_n\ket{n}, \quad  (n=0,1,2,\dots).
\end{eqnarray}
Then, the symmetrized correlation function $S(\omega)$ is described as
\begin{eqnarray}
    \label{eq:Sq}
    & & S(\omega)=\frac{\pi}{Z_q}\sum_{n,m}e^{-\beta\hbar\omega_{m}}
    |\langle n| \sigma_z | m \rangle|^2 \left[\delta(\omega-\omega_{nm})+\delta(\omega+\omega_{nm})\right],
\end{eqnarray}
where $Z_q={\rm tr}[e^{-\beta H_q}]$ and $\omega_{nm}=\omega_n-\omega_m$ is the transition frequency.
When switching on the system-bath coupling $\eta$, the delta function peaks in $S(\omega)$ are replaced by those with finite widths.
In general, their positions and widths vary when $\eta$ increases from zero.
However, in the weak-coupling regime ($\eta \ll 1$) these effects are so weak that the expression of $S(\omega)$ given in equation~(\ref{eq:Sq}) can be used approximately.

At low temperatures, $S(\omega)$ is determined mainly by transitions from the ground state.
Then, the symmetrized correlation function is approximated as
\begin{eqnarray}
    \label{eq:S_wc}
    & & S(\omega)\approx\pi\sum_n|\langle n |\sigma_z| 0\rangle |^2\left[\delta(\omega-\omega_{n0})+\delta(\omega+\omega_{n0})\right].
\end{eqnarray}
By substituting this expression into equation~(\ref{eq:conductance}), the thermal conductance is obtained by
\begin{eqnarray}
    \label{eq:conductance_wc}
    & & \kappa
    \approx\frac{\pi\alpha\gamma k_B}{16}\sum_{n}|\langle n|\sigma_z|0\rangle|^2\tilde{I}_{\rm eff}(\omega_{n0})\frac{(\beta\hbar\omega_{n0})^2}{\sinh(\beta\hbar\omega_{n0})}.
\end{eqnarray}
As a simple example, let us consider the condition $k_{\rm B}T \sim \hbar \omega_{10}$ and $\omega_{10} \ll \omega_{n0}$ for $n \ge 2$.
Then, the thermal conductance is obtained as
\begin{eqnarray}
    & & \kappa
    \approx\frac{\pi\alpha\gamma k_B}{16}\tilde{I}_{\rm eff}(\omega_{10})\frac{(\beta\hbar\omega_{10})^2}{\sinh(\beta\hbar\omega_{10})}.
\end{eqnarray}
In this example, the thermal conductance shows the Schottky-type temperature dependence~\cite{Segal2005PRL,Ruokola2011,Saito2013,Yamamoto2018NJP}; it has a peak at approximately $T\simeq \hbar\omega_{10}/k_B$ and shows exponential suppression ($\kappa\sim T^{-2}e^{-\hbar\omega_{10}/k_BT}$) for $T\ll\hbar\omega_{10}/k_B$.

If the transition energy to the second excited state, $\hbar\omega_{20}$, is also comparable to $k_{\rm B}T$, the temperature dependece of the thermal conductance is written as a sum of the two Schottky functions.
This implies that the thermal conductance may have {\it two peaks} as a function of the temperature.
We will discuss this possibility in section~\ref{sec:result} in details.
The present weak-coupling approximation is useful for qualitative discussion on the temperature dependence of the thermal conductance.
However, because the system-bath coupling actually affect the peak position and peak broadening, we need a more sophisticated approximation for quantitative discussion.
In the next section, we introduce an alternative approximation.

\section{NIBA}
\label{sec:NIBA}

In this section, we formulate the thermal conductance using the noninteracting-blip approximation (NIBA), which works well in a wide parameter region including the strong system-bath coupling regime \cite{Leggett1987,Weiss_text,Yamamoto2018NJP}.
The time evolution of the population $\Braket{\sigma_z(t)}$ is generally described by the generalized master equation
\begin{eqnarray}
    & & \frac{d\Braket{\sigma_z(t)}}{dt}
    = -\int_0^tdt'~K_z(t-t')\Braket{\sigma_z(t')}.
\end{eqnarray}
Here, the kernel $K_z(t-t')$ can describe non-Markivian dynamics of the population $\Braket{\sigma_z(t)}$.
In the NIBA, the non-local kernel takes a simple form,
\begin{eqnarray}
    & & K_z(t)=\Delta^2e^{-Q'(t)}\cos Q''(t),
\end{eqnarray}
where $Q(t)=Q'(t)+iQ''(t)$ is the complex bath correlation function, whose real and imaginary parts are given as
\begin{eqnarray}
    \label{eq:Q}
    & & Q'(t)=\int_0^\infty d\omega~\frac{I_{\rm eff}(\omega)}{\omega^2}\coth\left(\frac{\beta\hbar\omega}{2}\right)(1-\cos\omega t), \\
    & & Q''(t)=\int_0^\infty d\omega~\frac{I_{\rm eff}(\omega)}{\omega^2}\sin\omega t ,
\end{eqnarray}
respectively, where $I_{\rm eff}(\omega)=I_{{\rm eff},L}(\omega)+I_{{\rm eff},R}(\omega)$.
Using the effective spectral density given in equation~(\ref{eq:Ieff2}), the explicit expression{s} of $Q'(t)$ and $Q''(t)$ are calculated as~\cite{Nesi2007,Magazzu2019},
\begin{eqnarray}
    & & Q'(t)=\pi\alpha\left[\frac{2}{\hbar\beta}t-L\left(e^{-\Gamma t}\cos\bar{\Omega}t-1\right)-Ze^{-\Gamma t}\sin\bar{\Omega}t+Q'_{\rm Mats}(t)\right], \\
    & & Q''(t)=\pi\alpha\left[1-e^{-\Gamma t}\left(\cos\bar{\Omega}t-N\sin\bar{\Omega}t\right)\right],
\end{eqnarray}
where
\begin{eqnarray}
    & & \bar{\Omega}=\sqrt{\Omega^2-\Gamma^2},\\
    & & N=(\Omega^2-2\Gamma^2)/(2\Gamma\bar{\Omega}),\\
    & & L=\frac{N\sinh(\beta\hbar\bar{\Omega})-\sin(\beta\hbar\Gamma)}{\cosh(\beta\hbar\bar{\Omega})-\cos(\beta\hbar\Gamma)}, \\
    & & Z=\frac{\sinh(\beta\hbar\bar{\Omega})+N\sin(\beta\hbar\Gamma)}{\cosh(\beta\hbar\bar{\Omega})-\cos(\beta\hbar\Gamma)},
\end{eqnarray}
and $Q'_{\rm Mats}(t)$ is a series with respect to the Matsubara frequency $\nu_n=2\pi n/(\hbar\beta)$ defined as
\begin{eqnarray}
    & & Q'_{\rm Mats}(t)=\frac{4\Omega^4}{\hbar\beta}\sum_{n=1}^\infty\frac{1}{\nu_n}\frac{1-e^{-\nu_n t}}{(\Omega^2+\nu_n^2)^2-(2\Gamma\nu_n)^2}.
\end{eqnarray}

The symmetrized correlation function $S(\omega)$ can be related to the kernel as~\cite{Weiss_text}
\begin{eqnarray}
    & & S(\omega)={\rm Re}\left[\frac{2}{-i\omega+\hat{K}_z(-i\omega)}\right],
\end{eqnarray}
where $\hat{K}_z(\lambda)$ is the Laplace transform of $K_z(t)$.
By substituting this expression for the symmetrized correlation function into equation~(\ref{eq:conductance}), we can calculate the linear thermal conductance in the NIBA.
In general, the NIBA is justified (a) at arbitrary temperatures for the weak system-bath coupling or (b) for incoherent transport realized at the high temperature or the strong system-bath coupling~\cite{Weiss_text,Leggett1987}.
When the system-bath coupling is not weak, the NIBA fails at low temperatures where heat transport induced by virtual excitations is dominant~\cite{Yamamoto2018NJP}.
However, the deviation from the exact result remains small as far as the system-bath coupling is weak.

\section{Result}
\label{sec:result}

\begin{figure}[tb]
\centering
\includegraphics[width=0.6\columnwidth]{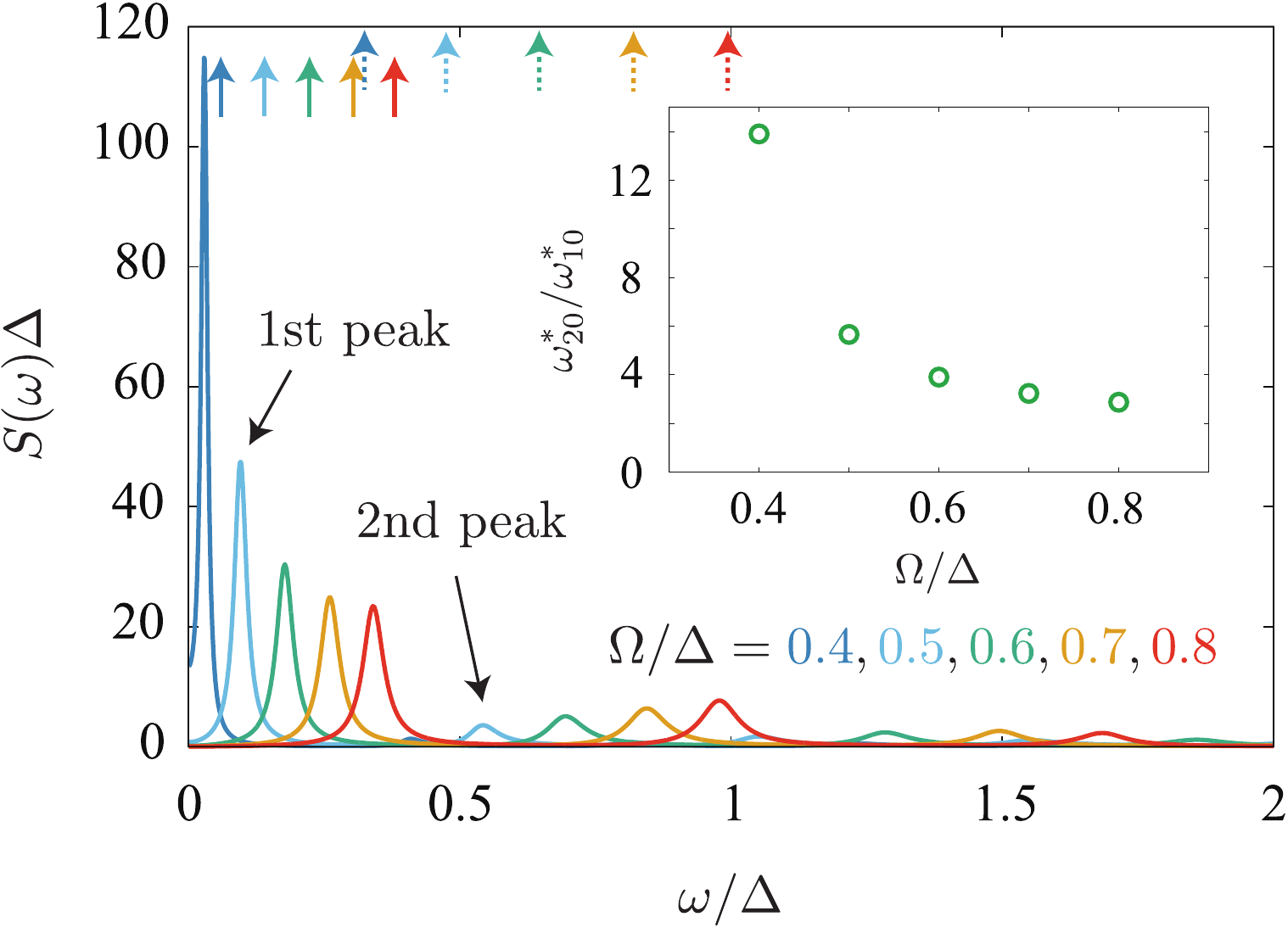}
\caption{\label{fig:Somega}
Symmetrized correlation function $S(\omega)$ as a function of the frequency for different $\Omega/\Delta$ at fixed other parameters $g=0.5\Delta,~\eta=0.01,~k_BT=0.01\hbar\Delta$.
$S(\omega)$ has several peaks and the peaks at the lower frequency are more dominant.
The black arrows indicate the first peak at $\omega_{10}^*=0.096\Delta$ and the second peak at $\omega_{20}^*=0.54\Delta$ for $\Omega=0.5\Delta$.
The solid and dotted arrows represent the lowest two bare transition frequencies, $\omega_{10}$ and $\omega_{20}$, respectively, which are obtained by the exact diagonalization of the quantum Rabi model~(\ref{eq:Hq}).
The inset represents a ratio between the frequencies of the first and second peaks, $\omega_{10}^*/\omega_{20}^*$, as a function of $\Omega/\Delta$.}
\end{figure}

In this section, we present the results calculated by the NIBA for $\Omega \lesssim \Delta$, where heat transport is expected to be non-trivial, as discussed in section~\ref{sec:ohmic}.

First, we show the symmetrized correlation function $S(\omega)$ in figure~\ref{fig:Somega} for $g=0.5\Delta$, $\eta=0.01$, and $k_{\rm B}T=0.01\hbar\Delta$.
As seen in the figure, $S(\omega)$ has several peaks corresponding to the transition frequencies of the quantum Rabi model as indicated by the approximation for the weak-coupling regime (see section~\ref{sec:weak_diss}).
However, the peak positions are shifted from the bare transition frequencies due to the system-bath coupling even for $\eta = 0.01$.
For later discussion, we define the frequency of the $n$-th peak in $S(\omega)$ as $\omega_{n0}^*$.
We note that $\omega_{n0}^*$ can be regarded as the renormalized transition frequencies corresponding to the bare transition frequencies $\omega_{n0}$ which are shown in figure~\ref{fig:Somega}.

\begin{figure}[tb]
\centering
\includegraphics[width=0.55\columnwidth]{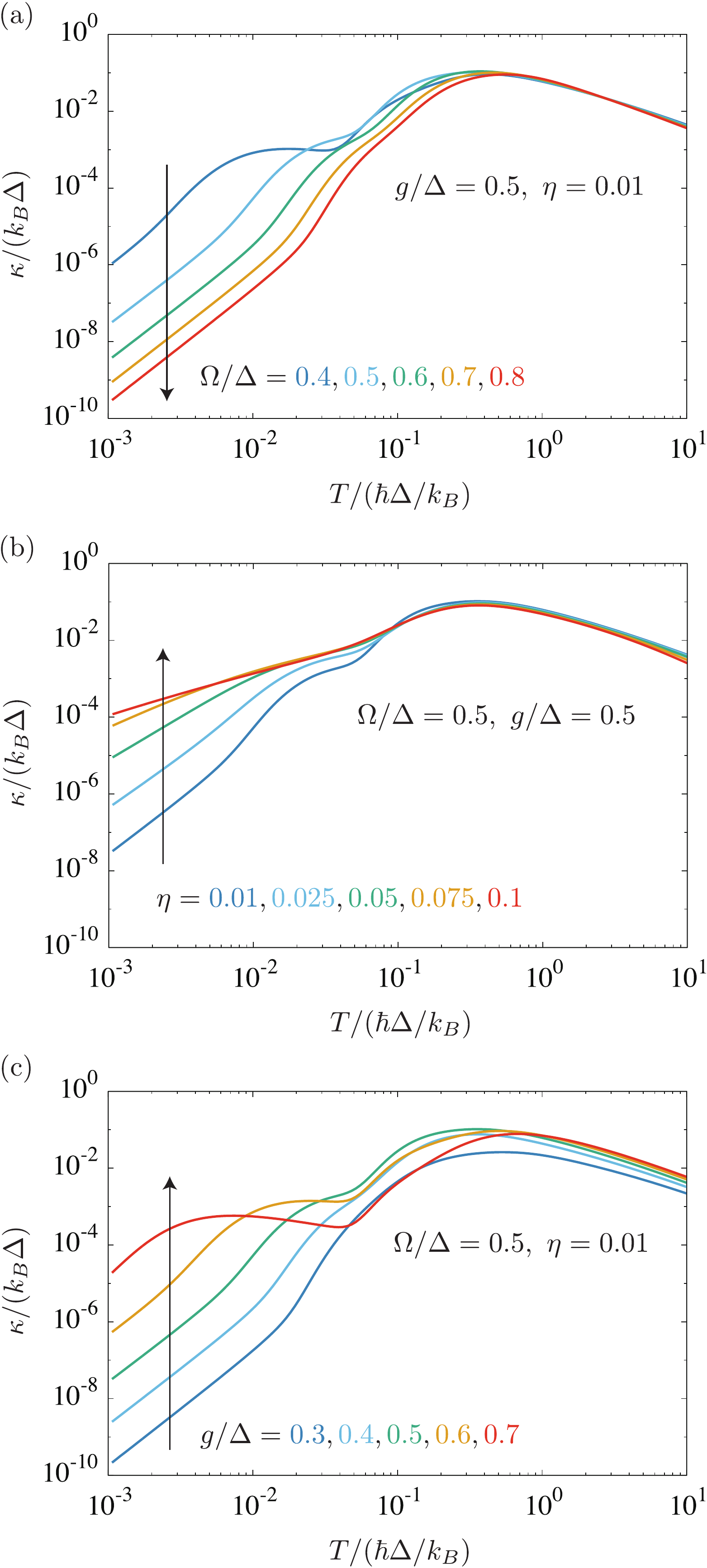}
\caption{\label{fig:kappa_vs_T}
Temperature dependence of the thermal conductance for typical parameters $\Omega/\Delta=0.5$, $g/\Delta=0.5$, and $\eta=0.01$.
When one varies (a) $\Omega/\Delta=0.4-0.8$, (b) $\eta=0.01-0.1$, and (c) $g/\Delta=0.3-0.7$, the thermal conductance considerably changes between the one- and two-peak structures.}
\end{figure}

Next, we show the temperature dependence of the thermal conductance in figure~\ref{fig:kappa_vs_T}~(a) for $g=0.5\Delta$ and $\eta=0.01$.
For $\Omega/\Delta = 0.4$, we observe a clear two-peak structure in the temperature dependence.
We also observe that the two-peak structure becomes less significant when $\Omega/\Delta$ is varied from 0.4 to 0.8.
This feature in the thermal conductance originates from the first and second peaks of $S(\omega)$, whose frequencies are given by $\omega_{10}^*$ and $\omega_{20}^*$ (see figure~\ref{fig:Somega}).
Then, its temperature dependence is qualitatively described by the sum of the two Schottky-type functions, $T^{-2}e^{-\hbar\omega_{10}^*/k_BT}$ and $T^{-2}e^{-\hbar\omega_{20}^*/k_BT}$, which leads to the double-peak structure.

The two-peak structure is a signature of the multiple levels characteristic of the quantum Rabi model and can be clearly observed only when the symmetrized correlation function has sharp well-separated peaks.
We show the temperature dependence of the thermal conductance for $\Omega=0.5\Delta$ and $g=0.5\Delta$ in figure~\ref{fig:kappa_vs_T}~(b).
When $\eta$ increases from 0.01 to 0.1, the peak at the low-temperature side becomes less significant.
This is because the peaks in $S(\omega)$ are broadened due to the system-bath coupling and are smoothed out after performing the integral in equation~(\ref{eq:conductance}).

Let us discuss the detailed condition for the appearance of the two-peak structure in the thermal conductance (\ref{eq:conductance}).
If the peaks in the symmetrized correlation function $S(\omega)$ are sufficiently sharp, the thermal conductance is approximately evaluated as
\begin{eqnarray}
    & & \kappa(T)
    \sim a_1\frac{e^{-1/\tilde{T}}}{(k_BT)^2} 
    + a_2\frac{e^{-(\omega_{20}^*/\omega_{10}^*)/\tilde{T}}}{(k_BT)^2},
    \label{eq:kappaTeff}
\end{eqnarray}
where $\tilde{T}=k_{\rm B}T/\hbar \omega_{10}^*$, and the prefactors are given as 
\begin{eqnarray}
    & & a_n=\frac{\pi\alpha\gamma k_B}{8} |\bra{n}\sigma_z\ket{0}|^2(\hbar\omega_{n0}^*)^2\tilde{I}_{\rm eff}(\omega_{n0}^*).
    \label{eq:kappaTeff2}
\end{eqnarray}
The two-peak structure in the temperature dependence of the thermal conductance appears when $\omega_{20}^*/\omega_{10}^* \gg 1$.
We note that although the ratio between $a_1$ and $a_2$ is also relevant to the appearance of the two peaks, it is a minor effect.
This condition can be confirmed by the comparison between figures~\ref{fig:Somega} and \ref{fig:kappa_vs_T}~(a).
When $\Omega/\Delta$ increases from 0.4 to 0.8, the ratio between the frequencies of the first and second peaks,  $\omega_{20}^*/\omega_{10}^*$, is reduced from 13.9 to 2.87 (see the inset in figure~\ref{fig:Somega}).
According to the reduction of $\omega_{20}^*/\omega_{10}^*$, the two-peak structure becomes less significant when $\Omega/\Delta$ increases, see figure~\ref{fig:kappa_vs_T}~(a).
This feature can be also confirmed by the dependence of the hybridization constant $g$.
Figure~\ref{fig:kappa_vs_T}~(c) shows the temperature dependence of the thermal conductance for $\Omega=0.5\Delta$ and $\eta=0.01$.
It is observed that when $g/\Delta$ increases from 0.3 to 0.7, the two-peak structure becomes more significant.
This is because the ratio $\omega_{20}^*/\omega_{10}^*$ is enlarged by the level repulsion when the hybridization $g$ increases.

\begin{figure}
\centering
\includegraphics[width=0.6\columnwidth]{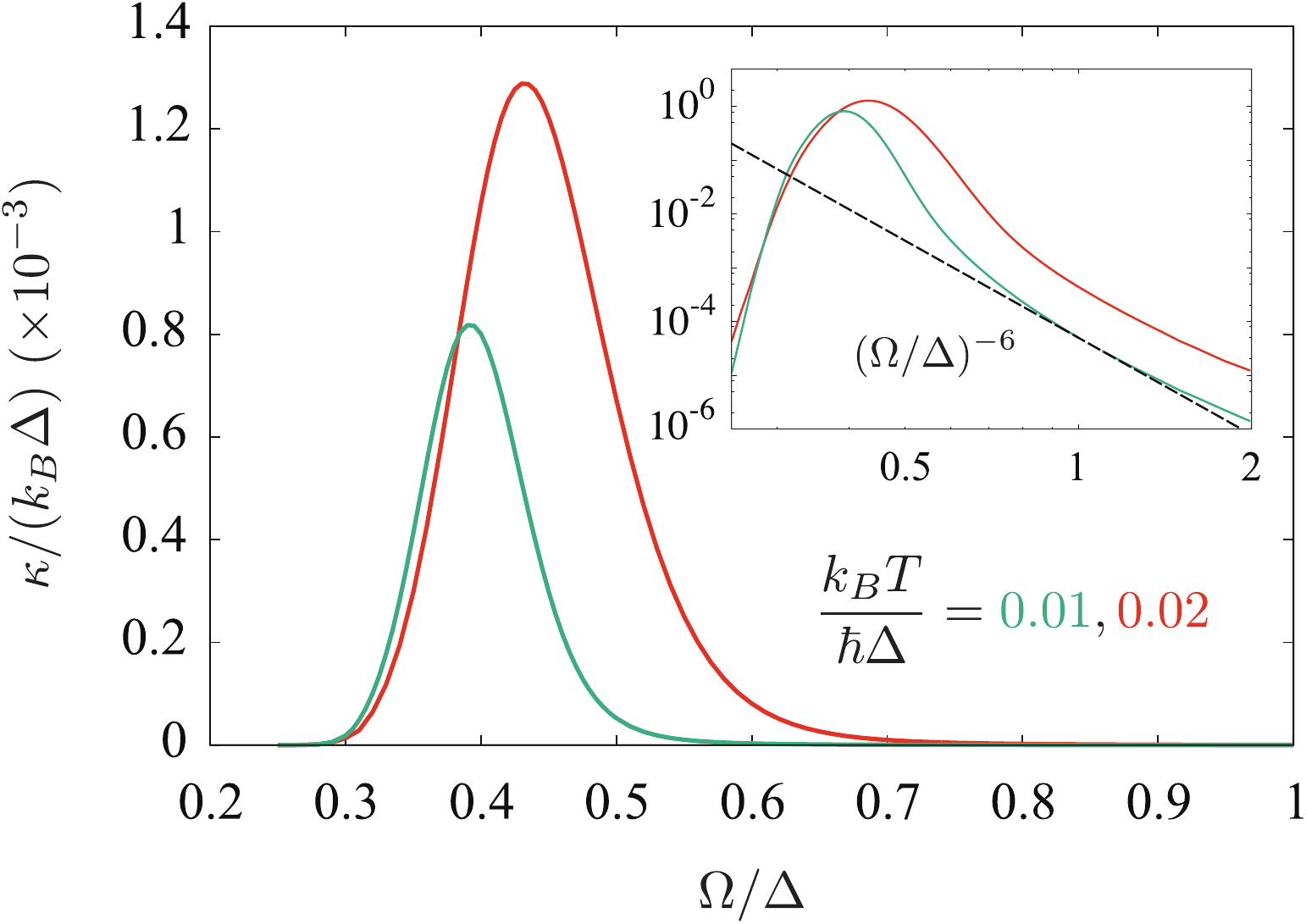}
\caption{\label{fig:kappa_vs_Omega}
Thermal conductance as a function of $\Omega/\Delta$ for $g/\Delta=0.5,~\eta=0.01$, and different temperatures $k_BT=0.01\hbar\Delta$ and $0.02\hbar\Delta$.
The inset is an enlarged view of the high-frequency side of the peak frequency.
The dashed line represents a guide for the power law $(\Omega/\Delta)^{-6}$.}
\end{figure}

When we choose the parameters for which the two-peak structure in $\kappa(T)$ is clearly observed, the thermal conductance is sensitive to the parameters, particularly to $\Omega/\Delta$.
In figure~\ref{fig:kappa_vs_Omega}, we show the thermal conductance as a function of $\Omega/\Delta$ for two different temperatures.
The thermal conductance has a sharp peak at $\Omega = \Omega_0$, where $\Omega_0 \approx0.40 \Delta$ and $0.44 \Delta$ for $k_B T = 0.01\hbar\Delta$ and $0.02\hbar\Delta$, respectively.
The thermal conductance decays exponentially in the small-$\Omega$ side of the peak, while it shows a power-law decay in the opposite side, whose exponent is $-6$ (see the inset in figure~\ref{fig:kappa_vs_Omega}).
The strong dependence of $\Omega/\Delta$ has an advantage for application of heat transistor or heat valves in mesoscopic heat devices.

Finally, we indicate that the power-law decay in $\kappa(\Omega)$ for large $\Omega/\Delta$ is a signature of the co-tunneling process in which heat transport is induced by the tunneling process via virtual excitations~\cite{Ruokola2011,Saito2013,Yamamoto2018NJP}.
At low temperatures ($k_BT\ll\hbar\Gamma,~\hbar\bar{\Omega},~\hbar\omega_{20}^*$), it is supposed the system behaves effectively like the Ohmic spin-boson model with a single parameter $\omega_{10}^*$ because the high-energy states are not relevant.
Furthermore, for $k_BT \lesssim 0.1\hbar\omega_{10}^*$, thermal excitation from the ground state to the first excitated state is strongly suppressed, and heat transport is governed by the co-tunneling process for which the thermal conductance is written as~\cite{Ruokola2011,Saito2013,Yamamoto2018NJP}
\begin{eqnarray}
    \label{eq:conductance_co}
    & & \kappa\approx\frac{\pi^3\gamma k_B^4}{30}\alpha^2\chi_0^2T^3,
\end{eqnarray}
where $\chi_0$ is the static susceptibility of the effective spin-boson model that is roughly approximated as $2/(\hbar\omega_{10}^*)$ for the weak coupling ($\alpha\ll1$).
Noting that $\alpha=\eta(4g/\Omega)^2$ and $\omega_{10}^*\propto\Omega$ for $\Omega \lesssim \Delta$, we obtain $\kappa(\Omega)\propto\Omega^{-6}$, which is consistent with the numerical result shown in the inset in figure~\ref{fig:kappa_vs_Omega}.

\section{Discussion}
\label{sec:discussion}

\subsection{Experimental feasibility}
\label{sec:ExperimentalFeasibility}

The most promising candidate for realizing a system described by the quantum Rabi model is a superconducting circuit, which can be fabricated by the current technology on the circuit QED.
The quantum Rabi model has been a fundamental model to study how strongly a small quantum system (superconducting qubit) can couple to a single bosonic mode (resonator)~\cite{Bosman2017, Forn-Diaz2010, Forn-Diaz2016, Niemczyk2010, Yoshihara2017, Yoshihara2018}, and its implementation technique accessible to the ultra-strong coupling regime has been developed.
In particular, a superconducting circuit composed of a flux qubit and an $LC$ resonator has realized a strong coupling of $g/\Delta\sim 10$~\cite{Yoshihara2017}, which is sufficiently large to examine characteristic heat transport predicted in this paper.

By coupling the resonators to superconducting waveguides, which play the role of the Ohmic heat bath, heat transport through a small quantum system described by the quantum Rabi model can be studied.
Indeed, a similar setup has been implemented experimentally using a transmon qubit and superconducting transmission lines, and the heat flow through them has been measured with high accuracy although it does not reach the ultra-strong coupling regime, $g/\Delta\sim 0.01$ in reference~\cite{Ronzani2018}.
We expect that our theoretical prediction can be examined in this experimental setup after improving the coupling $g$.

\subsection{Rotating wave approximation}
\label{sec:JC}

By applying the rotating wave approximation to the quantum Rabi model, we obtain the Jaynes-Cummings model.
While these two models give almost the same results in weak coupling regime ($g/\Delta \ll 1$), they give different results in the ultra-strong coupling regime ($g/\Delta \gtrsim 1$).
We note that the superconducting qubit-resonator system in the ultra-strong coupling regime is described not by the Jaynes-Cumming model but by the quantum Rabi model because the rotating wave approximation fails in that regime.
However, it would be instructive and useful to consider the Jaynes-Cummings model as a mathematical model or a model that may be realized in systems other than superconducting circuits. 
Since it is cumbersome to construct the NIBA for the Jaynes-Cummings model, we employ the approximation for the weak-coupling regime ($\eta\ll1$) discussed in section~\ref{sec:weak_diss}.

For the convenience of later discussion, we introduced new Pauli operators, $\tau_{x,y,z}$, defined as
\begin{eqnarray}
    & & \sigma_x = -\tau_z, \quad \sigma_y = \tau_y, \quad \sigma_z = \tau_x.
\end{eqnarray}
This corresponds to a rotation of the coordinate axes around the $y$ axis by $\pi/2$.
We also introduce the state vectors of the two-level system, $\ket{0}$ and $\ket{1}$, as eigenstates of $\tau_z (= \sigma_x)$ as
\begin{eqnarray}
    & & \tau_z \ket{0} = -\ket{0}, \quad 
    \tau_z \ket{1} = +\ket{1}.
\end{eqnarray}
For a decoupled system ($g=0$), $\ket{0}$ and $\ket{1}$ correspond to the ground and excited states of the two-level system, respectively.
Moreover, we define $\tau_+ = \ket{1}\bra{0}$ and $\tau_- = \ket{0}\bra{1}$.
Using these new operators and the identity 
$\tau_x = \tau_+ + \tau_-$, the Hamiltonian of the quantum Rabi model given in equation~(\ref{eq:Hq}) is rewritten as
\begin{eqnarray}
    & & H_{\rm q}
    =\frac{\hbar\Delta}{2}\tau_z+\sum_r\hbar\Omega_ra_r^\dagger a_r +\sum_r\hbar g_r (\tau_+ + \tau_-)(a_r+ a_r^\dagger).
\end{eqnarray}
After dropping the terms including $\tau_+ a^\dagger$ and $\tau_- a$ by the rotating wave approximation, we finally obtain the Jaynes-Cummings model as
\begin{eqnarray}
    & & H_{\rm JC}
    =\frac{\hbar\Delta}{2}\tau_z+\sum_r\hbar\Omega_ra_r^\dagger a_r +\sum_r\hbar g_r\left(\tau_+a_r+\tau_- a_r^\dagger\right).
\end{eqnarray}

For simplicity, we consider the symmetric case, $\Omega_L=\Omega_R=\Omega$ and $\eta_L=\eta_R=\eta$ hereafter.
We introduce new operators as follows:
\begin{eqnarray}
    & & A=\frac{a_L+a_R}{\sqrt{2}},\quad
    B=\frac{a_L-a_R}{\sqrt{2}}.
\end{eqnarray}
The operators, $A$ and $B$, correspond to symmetric and antisymmetric modes of the two harmonic oscillators.
The Hamiltonian is then rewritten as
\begin{eqnarray}
    & & H_{\rm JC}=H_{A}+H_{B}, \\
    & & H_A=\frac{\hbar\Delta}{2}\tau_z+\hbar\Omega A^\dagger A+\sqrt{2}\hbar g\left(\tau_+A+\tau_- A^\dagger\right), \\
    & & H_B=\hbar\Omega B^\dagger B.
\end{eqnarray}
Here, the antisymmetric mode is completely decoupled with the two-level system and does not contribute heat transport via the two-level system.
In the following, we consider only the Hamiltonian $H_A$.

The Jaynes-Cummings model conserves the total number of ``particles'', whose operator is defined as $\hat{N}=\tau_+\tau_-+A^\dagger A$.
Note that the quantum Rabi model does not have such a conserved quantity.
The basis of the Jaynes-Cummings model, $\ket{\tau} \otimes \ket{n}$, has an eigenvalue $N=\tau + n$, where $\tau$ ($=0,1$) and $n$ ($=0,1,2,\dots$) assign the states of the two-level system and the symmetric mode, respectively.
It is easy to show the ground state is given as $\ket{0} \otimes \ket{0}$.
For the excited states, there are two eigenstates labeled by $N$ ($\ge1$):
\begin{eqnarray}
    & & \ket{i;N} = a_{Ni} \ket{0} \otimes \ket{N}
    + b_{Ni} \ket{1} \otimes \ket{N-1}, \quad (i=1,2),
\end{eqnarray}
where $a_{Ni}$ and $b_{Ni}$ are quantum amplitudes.

At low temperatures, the symmetrized correlation function is determined by the transitions from the ground state dominantly.
Then, the symmetrized correlation function is written as
\begin{eqnarray}
    & & S_{\rm JC}(\omega)
    \approx\pi\sum_{i=1,2}|\Bra{i;1}\tau_x\Ket{0;0}|^2\delta(\omega-\omega_{i,1}+\omega_{0,0}),
    \label{eq:sjc}
\end{eqnarray}
where $\hbar\omega_{i,N}$ is an energy eigenvalue of $\ket{i;N}$.
Here, we have used the fact that the operator $\tau_x$ changes the eigenvalue $N$ only by one.
Thus, the existence of the conserved quantity strongly restricts the possible transitions from the ground state to the excited states.
However, the final expression (\ref{eq:sjc}) indicates that $S_{\rm JC}(\omega)$ has indeed two peaks at $\omega = \omega_{i,1}-\omega_{0,0}$ ($i=1,2$).
Following the discussion below equations~(\ref{eq:kappaTeff})-(\ref{eq:kappaTeff2}), we can conclude that the double-peak structure appears in the temperature dependence of the thermal conductance even in the Jaynes-Cummings model when $\omega_{20}/\omega_{10} \gg 1$.
Finally, we note that because the energy levels differ between these two models, particularly in the ultra-strong coupling regime~\cite{Kockum2019}, they give qualitatively different temperature dependences for the thermal conductance there.

\section{Conclusion}
\label{sec:summary}

We studied heat transport through a small quantum system described by the quantum Rabi model using the non-interacting blip approximation (NIBA).
The thermal conductance shows the two-peak temperature dependence due to the multiple levels in the quantum Rabi model.
In addition, we found that the thermal conductance is highly sensitive to the parameters, particularly the natural frequency of the harmonic oscillators in the strong hybridization region $\Omega \sim g \lesssim \Delta$.
This property is advantageous to the application of well-controllable nanoscale heat devices such as a heat transistor.

In general, heat transport through a quantum system with multiple levels may show the same sensitivity to the parameter when its lowest three energy levels are appropriately designed, as done in this work.
However, the feature of heat transport obtained in this paper is expected to be well examined in a physical system described by the quantum Rabi model because it is rather feasible to realize it in various physical systems.
The dissipative quantum Rabi model can also be regarded as the spin-boson model with the structured spectral density, as mentioned in section~\ref{sec:sb-model}.
Therefore, our result should be applicable to various physical systems (e.g., the chemical reactions in biomolecules~\cite{Garg1985} and atoms in cavities~\cite{Thorwart2000}), which are described by the spin-boson model with the structured spectral density.
The most promising candidate for realizing a system described by the quantum Rabi model is a superconducting circuit.
We hope that characteristic heat transport predicted in this work will be observed experimentally in the near future.

\section*{Acknowledgement}
This work was supported by Grant-in-Aid for JSPS Fellows Grant Number JP20J11318 (TY) and by JSPS KAKENHI Grant Number JP20K03831 (TK).

\appendix

\section{Mapping to the spin-boson model}
\label{app:mapping}

\begin{figure}
\centering
\includegraphics[width=0.7\columnwidth]{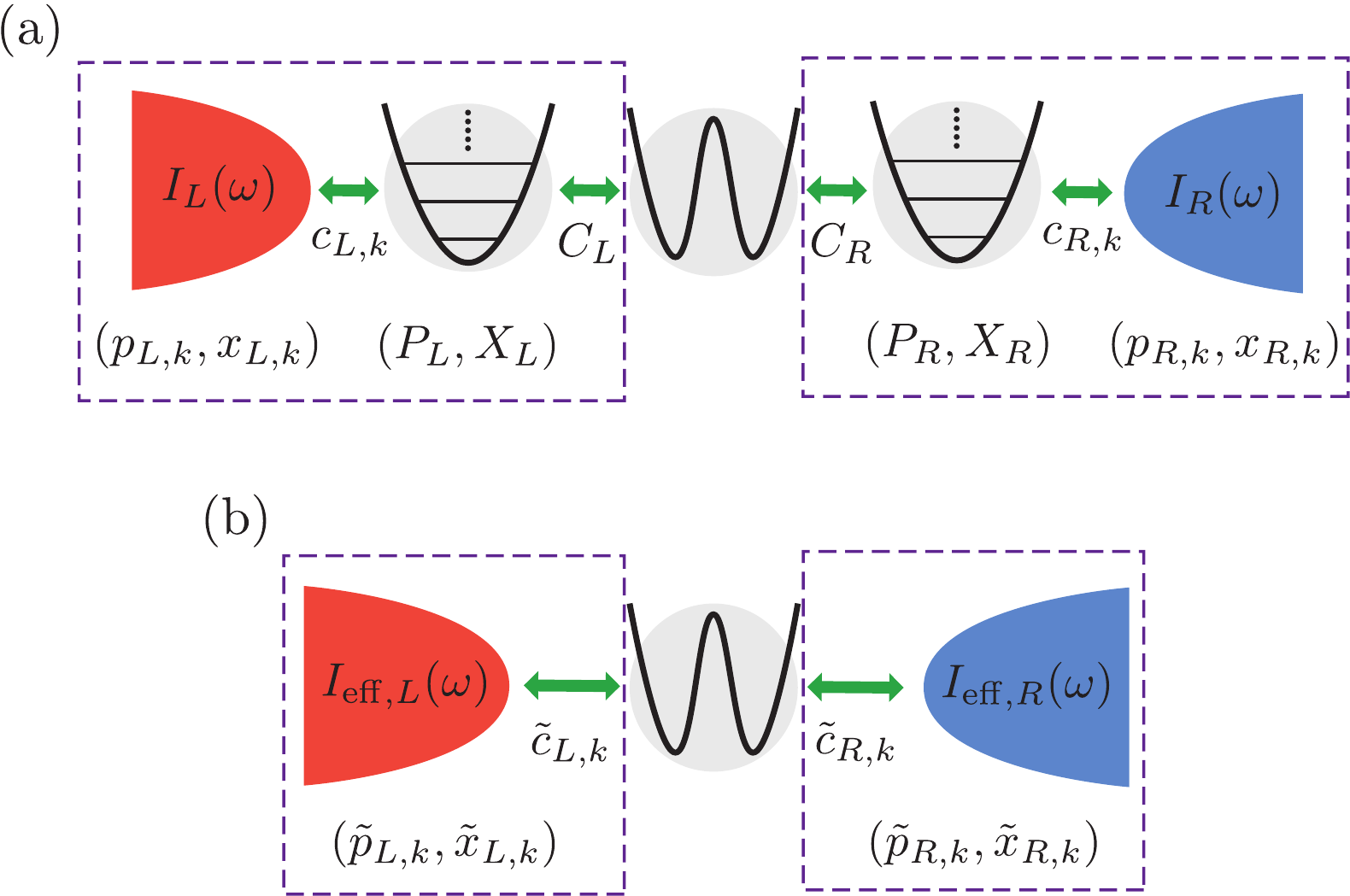}
\caption{\label{fig:mapping}
Schematic of the mapping used in this paper (see also figure~\ref{fig:system}).
(a) The original model (the dissipative quantum Rabi model). 
The momentum and position operators for the harmonic oscillators are denoted with $P_{r}$ and $X_{r}$ ($r=L,R$), respectively.
Similarly, those for the heat bath $r$ are denoted with $p_{r,k}$ and $x_{r,k}$, respectively.
The heat bath $r$ is characterized by the spectral density $I_r(\omega)$.
(b) The spin-boson model that is equivalent to the original model.
The composite system composed of the harmonic oscillator $r$ and the heat bath $r$ in the original model (indicated by the two dashed squares in the panel (a)) can be exactly mapped to the heat bath $r$ in the spin-boson model (the two dashed squares in the panel (b)).
For the heat bath $r$ in the spin-boson model, we introduce new momentum and position operators, $\tilde{p}_{r,k}$ and $\tilde{x}_{r,k}$, respectively, and the spectral density $I_{{\rm eff},r}(\omega)$.}
\end{figure}

In this appendix, we describe the detail of the mapping from the dissipative quantum Rabi model to the spin-boson model with the structured spectral density. 
We show a schematic of the mapping in figure~\ref{fig:mapping}.

The Hamiltonian of the original model (the dissipative quantum Rabi model) given in equations~(\ref{eq:Hq})-(\ref{eq:Hint}) is rewritten as
\begin{eqnarray}
    \label{eq:Hfull_class}
    & & H=H_{\rm S}+\sum_{r}\left(H_{{\rm osc},r}+H_{{\rm bath},r}\right) +\sum_{r}\left(H_{{\rm S}\hyphen{\rm osc},r}+H_{{\rm osc}\hyphen{\rm bath},r}\right), \\
    & & H_{\rm S}=-\frac{\hbar\Delta}{2}\sigma_x, \\
    & & H_{{\rm osc},r}=\frac{P_{r}^2}{2m_{r,0}}+\frac{1}{2}M_{r}\Omega_{r}^2 X_{r}^2, \\
    & & H_{{\rm bath},r}=\sum_k\left(\frac{p_{r,k}^2}{2m_{r,k}}+\frac{1}{2}m_{r,k}\omega_{r,k}^2x_{r,k}^2\right), \\    
    \label{eq:HI1}
    & & H_{{\rm S}\hyphen{\rm osc},r}=-\frac{1}{2} C_{r}\sigma_z X_{r}, \\
    \label{eq:H_bath_app}
    & & H_{{\rm osc}\hyphen{\rm bath},r}=-\sum_kc_{r,k} X_{r}x_{r,k}.
\end{eqnarray}
Here, $H_{\rm S}$, $H_{{\rm osc},r}$, and $H_{{\rm bath},r}$ describe the two-level system, the harmonic oscillator $r$ ($=L,R$), and the heat bath $r$, respectively.
For the harmonic oscillator $r$, the Hamiltonian is rewritten using a natural frequency $\Omega_r$, a mass $M_{r}$, a momentum operator $P_{r}$, and a position operator $X_{r}$.
The heat bath $r$ is described by a collection of the harmonic oscillators, where its Hamiltonian is similarly rewritten by using a natural frequency $\omega_{r,k}$, a mass $m_{r,k}$, a momentum operator $p_{r,k}$, and a position operator $x_{r,k}$.
The coupling between the two-level system and the harmonic oscillator is described by $H_{{\rm S}\hyphen{\rm osc},r}$, where $C_{r}$ denotes a coupling strength.
Similarly, the coupling between the harmonic oscillator and the heat bath is by $H_{{\rm osc}\hyphen{\rm bath},r}$, where $c_{r,k}$ denotes a coupling strength.
It is easy to check that the Hamiltonian~(\ref{eq:Hfull_class}) reproduces the original form given in the main text by introducing annihilation operators defined as 
\begin{eqnarray}
    & & a_r
    =\sqrt{\frac{M_{r}\Omega_{r}}{2\hbar}}\left(X_{r} +\frac{i}{M_{r}\Omega_{r}}P_{r}\right), \\
    & & b_{r,k}
    =\sqrt{\frac{m_{r,k}\omega_{r,k}}{2\hbar}}\left(x_{r,k}+\frac{i}{m_{r,k}\omega_{r,k}}p_{r,k}\right).
\end{eqnarray}
We note that the coupling strengths, $C_{r}$ and $c_{r,k}$, are related to the parameters used in the main text as
\begin{eqnarray}
    & & g_r = -\frac{C_{r}}{\sqrt{8\hbar M_{r}\Omega_{r}}},\quad
    \lambda_{r,k} =-\frac{c_{r,k}}{\sqrt{4M_{r} \Omega_{r} m_{r,k}\omega_{r,k}}},
    \label{eq:grexpression}
\end{eqnarray}
respectively.

Now, we consider a composite system composed of the harmonic oscillator and the heat bath as indicated by the dashed squares in figure~\ref{fig:mapping}~(a).
Its Hamiltonian is given as
\begin{eqnarray}
    & & H_{{\rm osc}+{\rm bath},r} = H_{{\rm osc},r}+H_{{\rm bath},r}+H_{{\rm osc}\hyphen{\rm bath},r}.
\end{eqnarray}
The key idea of the present mapping is that the Hamiltonian $H_{{\rm osc}+{\rm bath},r}$ only includes the harmonic oscillators, which are exactly solvable.
Indeed, the Hamiltonian $H_{{\rm osc}+{\rm bath},r}$ can be diagonalized by the canonical transformation~\cite{Fano1961,Kato2007} as
\begin{eqnarray}
    & & H_{{\rm osc}+{\rm bath},r}
    =\sum_k\left(\frac{\tilde{p}_{r,k}^2}{2\tilde{m}_{r,k}}+\frac{1}{2}\tilde{m}_{r,k}\tilde{\omega}_{r,k}^2\tilde{x}_{r,k}^2\right), 
\end{eqnarray}
where $\tilde{p}_{r,k}$ and $\tilde{x}_{r,k}$ are new momentum and position operators, respectively.
From this expression, we can regard the composite system as an effective heat bath for the spin-boson model, as indicated by the dashed squares in figure~\ref{fig:mapping}~(b).
The coupling between the effective heat bath and the two-level system is obtained by rewriting $H_{{\rm S}\hyphen{\rm osc},r}$ with the new operator $\tilde{x}_{r,k}$ as 
\begin{eqnarray}
    \label{eq:HI1_diag}
    & & \tilde{H}_{{\rm S}\hyphen{\rm bath},r}
    =-\frac{1}{2}\sigma_z\sum_k\tilde{c}_{r,k}\tilde{x}_{r,k},
\end{eqnarray}
where $\tilde{c}_{r,k}$ is a coupling strength between the two-level system and the effective heat bath.
We note that since the coupling between the two-level system and the heat bath takes the bilinear form after the canonical transformation, the resultant model becomes the spin-boson model.
Although we do not show an explicit form $\tilde{c}_{r,k}$, it is sufficient to use the relation
\begin{eqnarray}
    & & \sum_k\tilde{c}_{r,k}\tilde{x}_{r,k}=
    C_{r}X_{r}.
\end{eqnarray}
Using this relation, the spectral density of the effective heat bath can be written by
\begin{eqnarray}
    \label{eq:I-C}
    & & I_{{\rm eff},r}(\omega)=\frac{C_{r}^2}{\pi\hbar}{\rm Im}\left[G_r(\omega)\right].
\end{eqnarray}
Here, $G_r(\omega)$ is the Fourier transformation of a correlation function of the harmonic oscillator
\begin{eqnarray}
    & & G_r(t) =-\frac{1}{i\hbar}\theta(t)\Braket{\left[X_{r}(t),X_{r}(0)\right]}_{{\rm bath},r}, \\
    & & X_{r}(t) =e^{itH_{{\rm osc}+{\rm bath},r}/\hbar} X_r e^{-itH_{{\rm osc}+{\rm bath},r}/\hbar},
\end{eqnarray}
where $\Braket{\cdots}_{{\rm bath},r}$ indicates the thermal average with respect to $H_{{\rm osc}+{\rm bath},r}$.
Using the linear response theory, the Fourier transformation of the correlation function is obtained as~\cite{Weiss_text}
\begin{eqnarray}
    & & G_r(\omega)=\frac{1}{M_{r}}\frac{1}{\Omega_{r}^2-\omega^2-i\omega\gamma_r(\omega)},
\end{eqnarray}
with a friction kernel
\begin{eqnarray}
    & & \gamma_r(t)=\theta(t)\frac{1}{M_{r}}\sum_k\frac{c_{r,k}^2}{m_{r,k}\omega_{r,k}^2}\cos(\omega_{r,k}t).
\end{eqnarray}
Assuming the heat baths of the original model as the Ohmic baths, $I_r(\omega)=2\eta_{r}\omega$, the Fourier transformation of the friction kernel becomes a constant value, $\gamma_r(\omega)=2\pi\eta_r\Omega_{r}=2\Gamma_r$.
Therefore, the effective spectral density~(\ref{eq:I-C}) is calculated as
\begin{eqnarray}
    & & I_{{\rm eff},r}(\omega)=\frac{C_{r}^2}{\pi\hbar M_{r}}\frac{2\Gamma_r\omega}{(\Omega_{r}^2-\omega^2)^2+(2\Gamma_r\omega)^2}.
\end{eqnarray}
Using equation~(\ref{eq:grexpression}), we obtain equation~(\ref{eq:Ieff2}) in the main text.

\section*{References}
\bibliographystyle{iopart-num}
\bibliography{DHO}

\end{document}